\documentclass[prd, twocolumn, tightenlines, twoside, secnumarabic, superscriptaddress, preprintnumbers, nofootinbib, notitlepage]{revtex4-1}
\usepackage{graphicx}
\usepackage{epstopdf}
\usepackage{amsmath}
\usepackage{amsfonts}
\usepackage{amssymb}
\usepackage{appendix}
\usepackage{comment}
\usepackage{color}
\usepackage{slashed}
\usepackage{subfigure}
\usepackage{setspace}
\usepackage{footnote}
\usepackage{multirow}
\usepackage{mathrsfs}
\usepackage{hyperref}
\usepackage{cleveref}
\usepackage{marvosym}
\definecolor{niceblue}{rgb}{0.388235, 0.627451, 0.847059}
\definecolor{nicered}{rgb}{0.7,0.1,0.1}
\definecolor{nicegreen}{rgb}{0.1,0.5,0.1}
\hypersetup{colorlinks,citecolor= nicegreen,linkcolor= nicered}

\newcommand{\be}{\begin{equation}}
\newcommand{\ee}{\end{equation}}
\newcommand{\ba}{\begin{array}}
\newcommand{\ea}{\end{array}}
\newcommand{\bea}{\begin{eqnarray}}
\newcommand{\eea}{\end{eqnarray}}
\newcommand{\balg}{\begin{align}}
\newcommand{\ealg}{\end{align}}
\newcommand{\bit}{\begin{itemize}}
\newcommand{\eit}{\end{itemize}}
\newcommand{\trm}[1]{\textrm{#1}}

\newcommand{\globes}{\texttt{GLoBES}}

\newcommand{\Mpc}{\trm{\Mpc}}
\newcommand{\yr}{\trm{\yr}}
\newcommand{\eV}{\trm{\eV}}

\begin{document}

\singlespacing
\allowdisplaybreaks

\title{Reevaluating Reactor Antineutrino Anomalies with Updated Flux Predictions}

\author{Jeffrey M. Berryman}
\affiliation{Center for Neutrino Physics, Department of Physics, Virginia Tech, Blacksburg, VA 24061, USA}
\affiliation{Department of Physics and Astronomy, University of Kentucky, Lexington, KY 40506, USA}
\affiliation{Department of Physics, University of California, Berkeley, CA 94720, USA}

\author{Patrick Huber}
\affiliation{Center for Neutrino Physics, Department of Physics, Virginia Tech, Blacksburg, VA 24061, USA}


\begin{abstract}
Hints for the existence of a sterile neutrino at nuclear reactors are
reexamined using two updated predictions for the fluxes of
antineutrinos produced in fissions. These new predictions diverge in
their preference for the rate deficit anomaly, relative to previous
analyses, but the anomaly in the ratios of measured antineutrino
spectra persists. We comment on upcoming experiments and their ability
to probe the preferred region of the sterile-neutrino parameter space
in the electron neutrino disappearance channel.
\end{abstract}

\maketitle


\textbf{Introduction:} The evidence for the existence of additional
neutrino species, which we generically call ``sterile neutrinos,'' can
be broadly decomposed into three classes: anomalous $\overline{\nu}_e$
disappearance at reactors~\cite{Mention:2011rk}; anomalous $\nu_e$ disappearance at GALLEX
\cite{Hampel:1997fc} and SAGE \cite{Abdurashitov:1998ne}, i.e., the
gallium anomaly; and anomalous $\overline{\nu}_e$ appearance at LSND
\cite{Aguilar:2001ty} and MiniBooNE
\cite{Aguilar-Arevalo:2013pmq}. These individual pieces, however, do
not form a consistent whole. The reactor and gallium anomalies were
found to be compatible at only the 9\% level in
Ref.~\cite{Dentler:2018sju}. Moreover, the combination of these and
the absence of anomalous $\nu_\mu$/$\overline{\nu}_\mu$ disappearance
is incongruous with the LSND and MiniBooNE appearance anomalies. For
recent reviews on the status of light sterile neutrinos, see
Refs.~\cite{Diaz:2019fwt,Boser:2019rta}. 

It is natural to consider, then, why this picture breaks down. While
models of new physics have been proposed to explain these anomalies
\cite{Merle:2014eja,Bakhti:2015dca,Babu:2016fdt,Carena:2017qhd,
  Magill:2018jla,Doring:2018cob,Liao:2018mbg,Denton:2018dqq}, an
obvious starting point is to scrutinize each to establish how robust
it really is. Regarding the gallium anomaly, the $^{69,71}$Ga$(\nu_e,
e^-)^{69,71}$Ge cross sections have recently been reevaluated in
Ref.~\cite{Kostensalo:2019vmv} using an updated shell-model
calculation. The preference for a sterile neutrino is weaker than for
previous calculations, but the compatibility between the gallium and
reactor anomalies is improved to 16\% \cite{Kostensalo:2019vmv}.  On
the other hand, the LSND and MiniBooNE anomalies will be extensively
probed by the upcoming short-baseline program at Fermilab
\cite{Antonello:2015lea,Machado:2019oxb}.  In this letter, we focus on
the the electron neutrino \emph{disappearance} channel.

We have reevaluated the reactor antineutrino anomaly (RAA) using
updated predictions for the reactor antineutrino fluxes. The RAA is
comprised of two parts: a deficit in the observed number of
$\overline{\nu}_e$ interactions relative to predictions and the
existence of features in measured $\overline{\nu}_e$ spectral
ratios. We address both of these in what follows. Our results are
obtained with the publicly available software \globes \,
\cite{Huber:2004ka,Huber:2007ji}; the underlying data libraries used
to produce our results will be published in an accompanying software
paper~\cite{upcoming}.


\textbf{Updated Flux Predictions:} Over the past decade, the
Huber-Mueller (HM) $\overline{\nu}_e$ flux predictions
\cite{Mueller:2011nm,Huber:2011wv} have been the standard fluxes for
calculations at reactors. These are derived from measurements of the
aggregate $\beta$ spectra from the products of nuclear fissions
\cite{VonFeilitzsch:1982jw,Schreckenbach:1985ep,Hahn:1989zr}. These
$\beta$ spectra are then converted to obtain predictions for the
corresponding $\overline{\nu}_e$ spectra using virtual $\beta$
branches; see Refs.~\cite{Mueller:2011nm,Huber:2011wv} for
details. While the physics that enters into this procedure is broadly
well understood, we highlight two sources of systematic
uncertainty. (1) \emph{The accuracy of the underlying data.} It cannot be
excluded that there may be unaccounted-for systematics in the
measurements of aggregate $\beta$ spectra that bias the results. (2) \emph{The
theoretical understanding of the component $\beta$ decays.} The
conversion procedure assumes that all $\beta$ decays are of allowed
type. However, a large fraction -- up to 40\% -- of all decays are
so-called forbidden decays. This introduces large uncertainties related
to nuclear structure.

An alternative to the conversion method exists in the \emph{ab initio}
method. Here, one adds the spectra from every accessible $\beta$
branch of every fission fragment, with appropriate weights determined
by the cumulative fission fraction of each isotope, to determine the
overall $\beta$ spectrum. This method benefits from using
\emph{physical} $\beta$ branches in lieu of \emph{virtual} branches
and thus can avoid some of the approximations made in the conversion
method.  However, this technique is similarly reliant on experimental
data, both for the strengths of the individual $\beta$ branches and
for the fission yields, which have historically been lacking. In
recent years, the beta feeding data have been revisited and improved
using Total Absorption Gamma Spectroscopy (TAGS), see {\it
  e.g.}~\cite{Estienne:2019ujo}. Furthermore, \emph{ab initio}
calculations are also dependent on a theoretical understanding of
forbidden decays.

Two new antineutrino flux predictions have recently appeared in the
literature. An updated \emph{ab initio} calculation was recently
published in Ref.~\cite{Estienne:2019ujo}. There, it was found that
\emph{ab initio} predictions produce better agreement with the
$\overline{\nu}_e$ spectrum measured at Daya Bay than the HM predictions,
This is due, in large part, to a $\sim$10\% reduction in the flux from $^{235}$U, a feature
that is consistent with previous findings \cite{Gariazzo:2018mwd,Adey:2019ywk}.
The improvement largely stems from the improved $\beta$ spectrum feeding functions
obtains from TAGS measurements. However, these calculations do not go beyond the allowed
approximation.

An updated conversion method calculation has been presented in
Ref.~\cite{Hayen:2019eop}; we refer to this calculation as ``HKSS'' in
the remainder of this work. The significant improvement in HKSS is
that forbidden decays are including via nuclear shell model
calculations to describe the underlying microscopic physics, allowing
the authors to derive the relevant shape factors. The authors find an
unspecific enhancement of the antineutrino flux at energies above
4\,MeV relative to HM, mitigating somewhat the size of the infamous 5
MeV bump \cite{Ko:2016owz,Bak:2018ydk,DoubleChooz:2019qbj,
  Adey:2019ywk} and increasing the predicted antineutrino flux.

We have considered the impact of all three of these flux
predictions on the preference of the global reactor antineutrino
data set for a sterile neutrino. An important factor in these analyses
is the size of the theoretical uncertainties on the flux
predictions. The HKSS flux predictions are published with
uncertainties; see the appendix to Ref.~\cite{Hayen:2019eop}. In our calculations,
we use the uncertainties from their parameterized results. The \emph{ab initio}
fluxes, however, have no stated uncertainties. In the absence of a more
compelling option, we assign the fractional uncertainties on the HM
predictions to the \emph{ab initio} predictions in our analysis. This
is an optimistic assignment of uncertainties; we will argue, however,
 that this does not affect the conclusions of this work.


\textbf{The Rate Anomaly:} We begin with combined analyses of the
inverse beta decay (IBD) event rates measured at the short-baseline
experiments at Bugey \cite{Declais:1994ma,Declais:1994su}, G\"{o}sgen
\cite{Zacek:1986cu}, ILL \cite{Kwon:1981ua,ILLupdate}, Krasnoyarsk
\cite{Vidyakin:1987ue,Vidyakin:1994ut,Kozlov:1999ct}, Nucifer
\cite{Boireau:2015dda}, Savannah River \cite{Greenwood:1996pb} and
Rovno \cite{Afonin:1988gx,Kuvshinnikov:1990ry}. Additionally, we
analyze Chooz \cite{Apollonio:2002gd}, Double Chooz
\cite{DoubleChooz:2019qbj} and Palo Verde \cite{Boehm:2001ik,GGratta}
at medium baselines, as well as fuel evolution results from Daya Bay
\cite{An:2017osx} and RENO \cite{RENO:2018pwo,SBKim}. We highlight the
salient features of our analysis here; see Ref.~\cite{upcoming} for
more details.

Our analysis is constructed using ratios of the IBD rates measured at
these experiments relative to the three-neutrino predictions for the
three reactor antineutrino flux models mentioned previously. We use
\globes \, to calculate the total event rate at each experiment as a
function of two sterile-neutrino parameters -- the effective mixing
angle $\sin^2 2\theta_{ee}$ and the mass-squared splitting $\Delta
m_{41}^2$. For short-baseline experiments, we use the two-flavor
approximation for the survival probability,
\begin{equation}
\label{eq:twoflavor}
P_{ee} \approx 1 - \sin^2 2\theta_{ee} \sin^2 \left( \frac{\Delta m_{41}^2 L}{4E_\nu} \right).
\end{equation}
For the medium-baseline experiments, we use the full four-flavor
oscillation formalism with the best-fit values for the usual
three-neutrino oscillation parameters from
Ref.~\cite{Esteban:2018azc}.\footnote{We assume that the existence of
 a sterile neutrino has not caused any of these parameters to be mismeasured.}

\begin{figure}[!t]
\includegraphics[width=\linewidth]{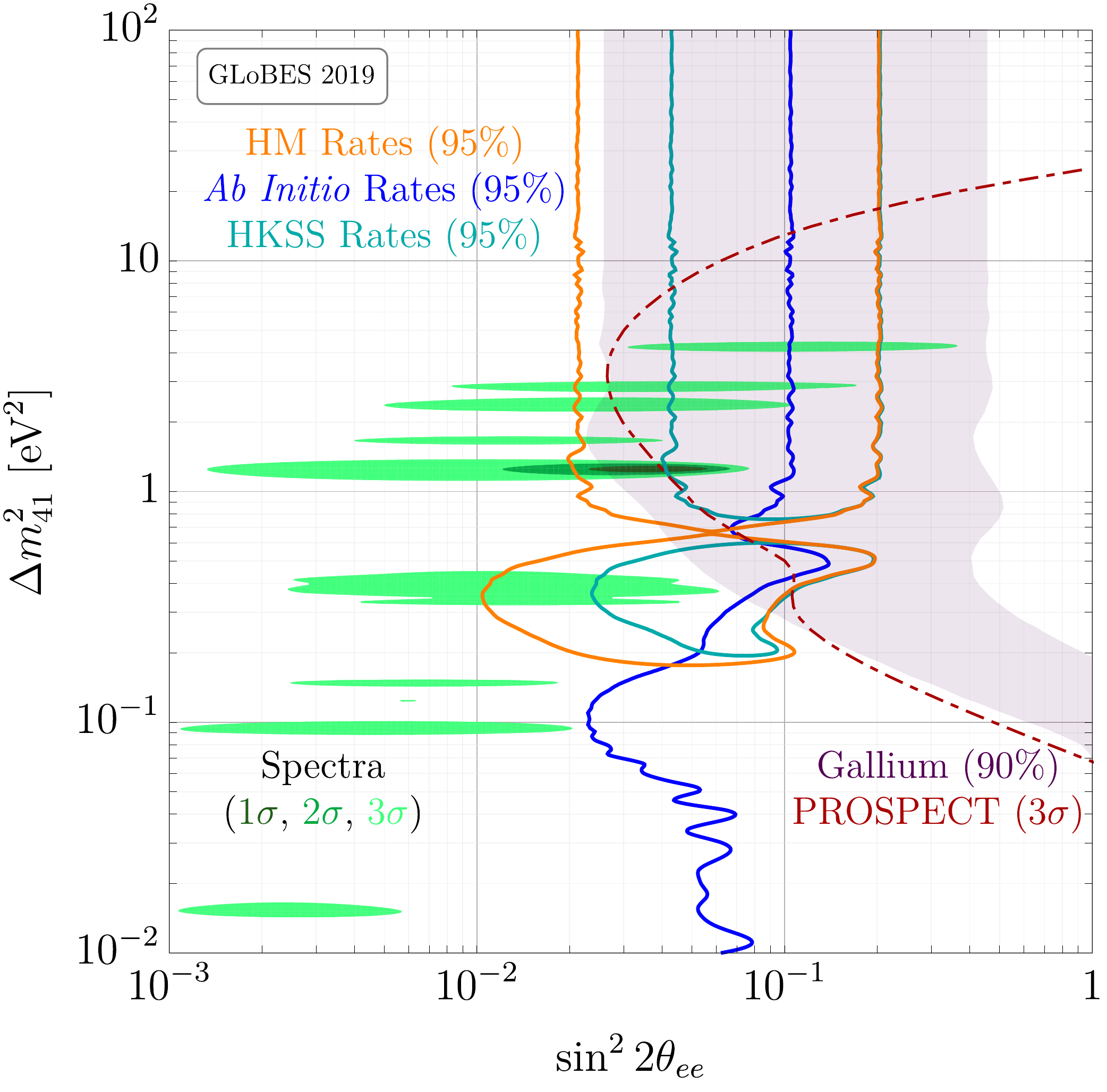}
\caption{The 95\% C.L. contours from IBD rate measurements 
using the HM (orange), \emph{ab initio} (blue) and HKSS 
(dark cyan) flux predictions. The regions
  preferred by reactor antineutrino spectra at
  $1\sigma$, $2\sigma$ and $3\sigma$ are shown in light, medium and
  dark green, respectively. We show the 90\% C.L. region preferred by the
  gallium anomaly \cite{Kostensalo:2019vmv}, for comparison. The red,
  dot-dashed curve shows the $3\sigma$ sensitivity of PROSPECT \cite{Ashenfelter:2015uxt}
  assuming three years of operation.}
\label{fig:results}
\end{figure}

The differences between the experimental and predicted ratios are combined into a global $\chi^2$ function, 
\begin{align}
\chi^2 = & \, (\vec{R}_{\rm exp} - \vec{R}_{\rm pred})^T \cdot
V^{-1}_{\rm exp} \cdot (\vec{R}_{\rm exp} - \vec{R}_{\rm pred})
\nonumber \\ & + \vec{\xi}^T \cdot V^{-1}_{\rm th} \cdot \vec{\xi},
\end{align}
where $\vec{R}_{\rm exp}$ is the vector of experimental ratios,
$\vec{R}_{\rm pred} = \vec{R}_{\rm pred}(\sin^2 2\theta_{ee}, \Delta
m_{41}^2, \vec{\xi} )$ is the vector of predicted ratios and
$\vec{\xi}$ is a vector of nuisance parameters describing the
normalization uncertainties on the isotopic flux predictions -- one
each for $^{235}$U, $^{238}$U, $^{239}$Pu and $^{241}$Pu. Further,
$V_{\rm exp}$ is the covariance matrix describing experimental
uncertainties, including correlations, and $V_{\rm th}$ is the
covariance matrix for $\vec{\xi}$. We minimize over the $\vec\xi$ for
each point in $\sin^2 2\theta_{ee}$--$\Delta m_{41}^2$ parameter
space.

We cross-check our results with the HM flux model against the
\emph{rate} results in
Refs.~\cite{Mention:2011rk,Gariazzo:2017fdh,Giunti:2017yid,Dentler:2017tkw,Dentler:2018sju,Giunti:2019qlt}. We
find general good agreement; the resulting
95\% C.L. curve is shown in orange in Fig.~\ref{fig:results}. The HM
fluxes are then replaced in favor of the \emph{ab initio} and HKSS
fluxes and the analysis is repeated; the resulting 95\% C.L. curves
are shown, respectively, in blue and dark cyan in
Fig.~\ref{fig:results}. For context, we show the region
preferred by the gallium anomaly at 90\% C.L. \cite{Kostensalo:2019vmv}
in shaded purple.

The updated flux models diverge, relative to the HM fluxes, in their
preference for a sterile neutrino. On one hand, the \emph{ab initio}
fluxes indicate a much \emph{weaker} preference for a sterile
neutrino; these fluxes prefer nonzero mixing at $<1\sigma$. This can
be largely attributed to the reduced total flux from $^{235}$U
fissions relative to the HM predictions, as mentioned above. Further,
recall that assigning the HM uncertainties to the \emph{ab initio}
fluxes underestimates the true theoretical uncertainty. A more
realistic error budget would further degrade the
preference for a sterile neutrino. On the other hand, the HKSS
predictions result in \emph{stronger} evidence for a sterile neutrino:
recalculating the shape factor accounting for forbidden decays results
in an increased expected IBD rate, implying larger experimental
deficits. Relevant statistics for these analyses are compiled in Table
\ref{tab:Stats}.

We conclude this discussion by underscoring that the diverging
preference for a sterile neutrino between the \emph{ab initio} and
HKSS flux predictions highlights the need to reappraise the data
underpinning these predictions. As of present, improved TAGS measurements
in the \emph{ab initio} model and the more complete treatment of forbidden
decays in HKSS modify the total predicted rate to roughly the same degree
but with opposite signs. Concerns about vastly increased uncertainties
from first-forbidden decays \cite{Hayes:2013wra} seem not to be borne out in the detailed analysis
in HKSS. That said, these conclusions can only be solidified with the collection
of more and improved data.

\begin{table}[!t]
\centering
\begin{tabular}{|c||c|c|c|c|c|}\hline
Analysis & $\chi^2_{3\nu}$ & $\chi^2_{\rm min}$ & $n_{\rm data}$ &  $p$ & $n\sigma$ \\ \hline \hline
HM Rates & 41.4 & 33.5 & 40 & $2.0\times10^{-2}$ & 2.3 \\ \hline
\emph{Ab Initio} Rates & 39.2 & 37.0 & 40 & 0.34 & 0.95 \\ \hline
HKSS Rates & 58.1 & 47.5 & 40 & $5.0\times10^{-3}$ & 2.8 \\ \hline \hline
Spectra & 184.9 & 172.2 & 212 & $1.8 \times 10^{-3}$ & 3.1 \\ \hline
DANSS + NEOS & 98.9 & 84.7 & 84 & $8.1 \times 10^{-4}$ & 3.3 \\ \hline
\end{tabular}
\caption{A summary of relevant statistics in our analyses. We show
  $\chi^2$ for $\sin^2 2\theta_{ee} = 0$, $\chi^2_{3\nu}$, and the
  minimum value of $\chi^2$ over the sterile neutrino parameter space,
  $\chi^2_{\rm min}$. We also tabulate the number of data points for
  each analysis, $n_{\rm data}$, the $p$-value at which three-neutrino
  mixing can be excluded and the number of $\sigma$ corresponding to
  that $p$-value.}
\label{tab:Stats}
\end{table}


\textbf{The Spectral Anomaly:} We shift our attention to the reactor
$\overline{\nu}_e$ energy spectra measured at Bugey
\cite{Declais:1994su}, DANSS \cite{Alekseev:2018efk}, Daya Bay
\cite{Adey:2018zwh}, Double Chooz \cite{DoubleChooz:2019qbj}, NEOS
\cite{Ko:2016owz} and RENO \cite{Bak:2018ydk}. With the exception of
NEOS, each of these experiments measures the $\overline{\nu}_e$
spectrum at multiple positions and publishes ratios of these
spectra. The benefit of such ratios is that the dependence on the
reactor flux model largely cancels, mitigating theoretical
uncertainties. The NEOS collaboration presents their spectrum as a
ratio with respect to the spectrum measured at Daya Bay in
Ref.~\cite{An:2016srz}, which introduces mild flux model dependence
into the analysis; see Ref.~\cite{upcoming} for details.

PROSPECT \cite{Ashenfelter:2018iov} and STEREO
\cite{Almazan:2018wln,Bernard:2019jli} have also produced constraints
in the last few years. Given that these experiments are still
collecting data and that only limited information on how
to include them in a global fit is available, we choose not to include them
here. We discuss their expected impact below.

The two-flavor approximation in Eq.~\eqref{eq:twoflavor} is used for
Bugey, DANSS and NEOS, but we use the full four-neutrino framework for
Daya Bay, Double Chooz and RENO. These spectral ratios are combined in
a single $\chi^2$ function of the form
\begin{equation}
\chi^2 = \sum_A (\vec{S}^A_{\rm exp} - \vec{S}^A_{\rm pred})^T \cdot \left(V_A\right)^{-1} \cdot (\vec{S}^A_{\rm exp} - \vec{S}^A_{\rm pred}),
\end{equation}
where $A$ indexes the experiments, $\vec{S}^A_{\rm exp}$ is the
experimental spectral ratio and $\vec{S}^A_{\rm pred} = \vec{S}^A_{\rm
  pred}(\sin^2 2\theta_{ee}, \Delta m_{41}^2)$ is the predicted
spectral ratio. Each experiment has its own covariance matrix $V_A$
that includes both experimental and theoretical uncertainties. In
principle, all experiments are correlated through the theoretical
uncertainties. Practically speaking, these correlations are
negligible.

The $\chi^2$ is calculated at each point in the $\sin^2
2\theta_{ee}$--$\Delta m_{41}^2$ parameter space; the results are
shown in Fig.~\ref{fig:results}. The $1\sigma$, 2$\sigma$ and
3$\sigma$ preferred regions are shown in dark, medium and light green,
respectively, and are consistent with similar results in
Refs.~\cite{Dentler:2017tkw,Gariazzo:2018mwd,Dentler:2018sju}. The
sensitivity is primarily driven by DANSS; the total evidence for a
sterile neutrino is $3.1\sigma$. It is noteworthy that NEOS and DANSS
point to the same $\Delta m_{41}^2$ despite their baselines differing
by a factor of two. Relevant statistics are compiled in the last line
of Tab. \ref{tab:Stats}.

We do not combine our rate and spectral analyses; there are nontrivial
correlations between the rate measurements at Bugey, Daya Bay, Double
Chooz and RENO and the corresponding spectral measurements that would
need to be taken into account. However, one can infer from
Fig.~\ref{fig:results} that the spectral analysis is consistent with
the \emph{ab initio} analysis; the latter shows weak preference for a
sterile neutrino, so consistency is essentially guaranteed. However,
one can also infer that the tension between the spectral and HKSS
analysis is greater than with the HM analysis. In this way, too, we see the
\emph{ab initio} and HKSS analyses diverge.


\textbf{Future Experiments:} It is useful and imperative to consider
how this parameter space can be probed in the near term, given the
uncertainty surrounding analyses of the rates but the apparent
robustness of spectral measurements. We consider only experiments
searching for $\nu_e$/$\overline{\nu}_e$ disappearance; for
discussions on the future of $\nu_e$/$\overline{\nu}_e$ appearance and
$\nu_\mu$/$\overline{\nu}_\mu$ appearance/disappearance, see
Refs.~\cite{Diaz:2019fwt,Boser:2019rta}.

We begin with PROSPECT and STEREO, which have produced early results
\cite{Ashenfelter:2018iov,Almazan:2018wln,Bernard:2019jli}, but not,
at present, final analyses. These experiments were designed in the
first half of the decade to conclusively probe the RAA as
presented in Ref.~\cite{Mention:2011rk}; early results indicate that
they will achieve this. However, since these experiments were
conceived, reactor spectrum experiments have shifted the preferred
sterile neutrino parameters to smaller mixing angles than previously
indicated.

We use PROSPECT as proxy to study how well current-generation reactor
can probe the regions preferred by the four global analyses presented
here. The expected $3\sigma$ sensitivity for three years of operation
is shown in dot-dashed dark red in Fig.~\ref{fig:results}
\cite{Ashenfelter:2015uxt}. This sensitivity represent a prediction of
how a null result from PROSPECT, {\it i.e.} no evidence for
oscillation, would constrain the parameter space. The question then
is, how much of the currently allowed regions would survive? To
quantify this, we calculate the difference between the three-neutrino
$\chi^2$, $\chi^2_{3\nu}$, and the minimum four-neutrino $\chi^2$ for
fixed $\Delta m_{41}^2$, $\chi^2_{\rm min}(\Delta m_{41}^2)$ -- a
measure of the preference the data show for oscillation at a given
$\Delta m_{41}^2$ -- for each analysis presented here.  Specifically,
we consider how a null result from PROSPECT would reduce this quantity
and hence, how much the allowed parameter space is reduced. 

\begin{figure}[!t]
\includegraphics[width=\linewidth]{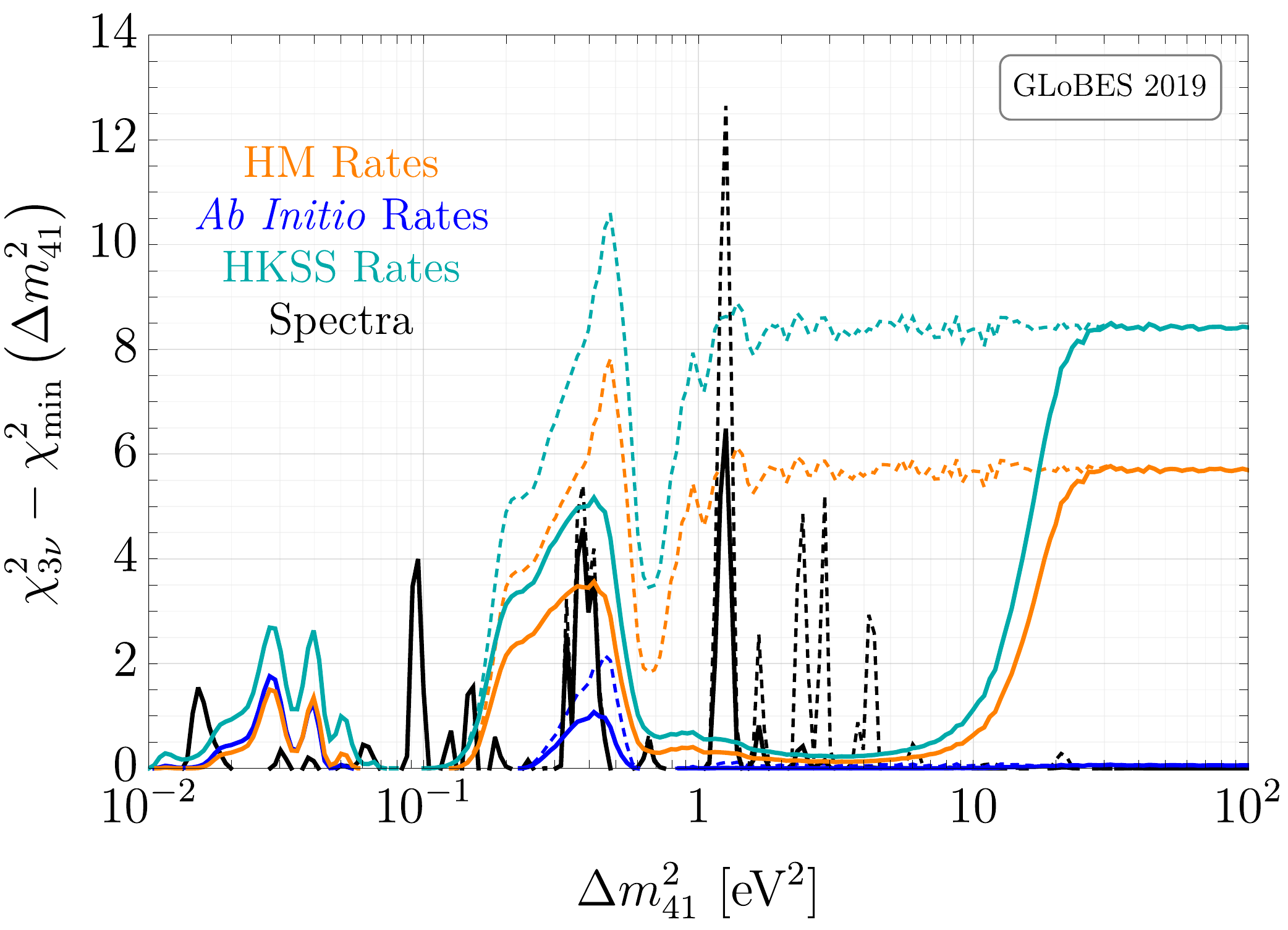}
\caption{The difference between the three-neutrino $\chi^2$,
  $\chi^2_{3\nu}$, and the minimum four-neutrino $\chi^2$ for a fixed
  $\Delta m_{41}^2$, $\chi^2_{\rm min}(\Delta m_{41}^2)$. The dashed
  lines are for the four global analyses presented here: HM rates
  (orange), \emph{ab initio} rates (blue), HKSS rates (dark cyan) and
  spectra (black); the solid lines show the remaining preference for
  oscillation after the inclusion of a hypothetical null result from
  PROSPECT.}
\label{fig:PROSPECT}
\end{figure}

The results are shown in Fig.~\ref{fig:PROSPECT}. The dashed curves
show $\chi^2_{3\nu} - \chi^2_{\rm min}(\Delta m_{41}^2)$ for the HM
rates (orange), \emph{ab initio} rates (blue), HKSS rates (dark cyan)
and spectra (black) analyses with no contribution from PROSPECT. The
solid curves are similar, except PROSPECT's sensitivity has been
folded in. In regions where the difference between the dashed and
solid curves of a given color is large, PROSPECT reduces the allowed
parameter space significantly.  PROSPECT is effective at probing the
HM and HKSS rate analyses in the region $\Delta m_{41}^2 \sim 1-10$
eV$^2$; this is precisely the region of parameter space for which it
was designed. Conversely, PROSPECT does little to challenge the
\emph{ab initio} rate analysis, since this produces weak preference
for a sterile neutrino in the first place.

Most interesting is its sensitivity to the spectral
anomaly. Measurements of spectral ratios are susceptible to
statistical fluctuations that ensure a best-fit point at nonzero
mixing; see Ref.~\cite{Agostini:2019jup} for more discussion. The
precise $1\sigma$ and $2\sigma$ preferred regions might be the result
of such fluctuations, but the $3\sigma$ preferred region is less
likely to be influenced by these. Consequently, we advocate taking a
broader view of the preferred parameter space when considering the
impact of PROSPECT on the spectral anomaly. PROSPECT is strongest in
probing the region $\Delta m_{41}^2 \sim 1-5$ eV$^2$, similar to the
HM and HKSS analyses.  However, it only has modest power to probe the
region $\Delta m_{41}^2 \sim 0.1- 0.5$, where the preference for a
sterile neutrino is also nontrivial. Overall, a $2.4\sigma$ preference
for oscillation would remain even after a null result from PROSPECT.

IsoDAR~\cite{Alonso:2017fci}, which is a proposal to use a
high-intensity $\overline{\nu}_e$ source from $\beta$ decays of
$^8$Li, has a expected $5\sigma$ sensitivity~\cite{Diaz:2019fwt} that
suggests it has the potential to emphatically confirm or refute the
sterile-neutrino interpretation of the RAA. There also exists a
burgeoning program of experiments searching for coherent elastic
neutrino-nucleus scattering \cite{Wong:2008vk,Akimov:2012aya,
  Gutlein:2014gma,Belov:2015ufh,Aguilar-Arevalo:2016qen,Agnolet:2016zir,Billard:2016giu,
  Strauss:2017cuu,Akimov:2017ade,CONUStalk} that may be able to probe
the anomaly \cite{Canas:2017umu, Blanco:2019vyp} at high significance,
though this process has not yet been observed at a nuclear reactor.


\textbf{Conclusions:} We have reanalyzed the global reactor
$\overline{\nu}_e$ data set using three reactor antineutrino flux
predictions. Relative to the traditional HM predictions, the two new
calculations result in diverging evidence for a sterile neutrino when
total IBD rate measurements are considered -- the \emph{ab initio}
calculation decreases the significance from $2.3\sigma$ to
$0.95\sigma$, whereas the HKSS calculation increases the significance
to $2.8 \sigma$. However, the spectral anomaly is robust with respect
to varying the flux model and is found to persist at the $3.1\sigma$
level. We have briefly reviewed the capabilities of current searches
for sterile neutrinos -- PROSPECT, in particular -- to cover the
parameter space preferred by each of these four analyses and find that a 
significant fraction of parameter space would remain even for a null
result in those searches.

We are aware of the optics of the current situation: the region of
parameter space preferred by the spectral anomaly lies \emph{just}
beyond the reach of past and current experiments. However, this
feature is shared with genuine discoveries; for example, $\theta_{13}$
was just beyond the sensitivity of the Chooz experiment. We have shown
that null results from current reactor experiments would leave a
significant fraction of the currently favored parameter space
unexplored.  Furthermore, the data in the electron disappearance
channel are self-consistent, irrespective of the flux model
used. Proclamations of the demise of the light sterile neutrino are,
therefore, premature.

Given the current evidence for the existence of light sterile
neutrinos, it is crucial that the next generation of oscillation
experiments includes an effective strategy for probing the
sterile neutrino hypothesis in the electron neutrino disappearance
channel.


\textbf{Acknowledgments:} We thank Giorgio Gratta (Palo Verde);
Soo-Bong Kim (RENO); David Lhuillier (STEREO); and Karsten Heeger,
Bryce Littlejohn and Pranava Teja Surukuchi (PROSPECT) for providing
data and useful discussions. We also thank Muriel Fallot for providing
the \emph{ab initio} fluxes in machine-readable format and Leendert
Hayen for information on the HKSS model.  JMB thanks the Fermilab
Theory Division for their hospitality during the completion of this
work.  This work is supported by DOE Office of Science
awards~DE-SC0018327 and~DE-SC0020262. The work of JMB is also
supported by NSF Grant~PHY-1630782 and by Heising-Simons Foundation
Grant~2017-228.
  

\bibliographystyle{apsrev-title}
\bibliography{references}{}

\begin{thebibliography}{79}
\expandafter\ifx\csname natexlab\endcsname\relax\def\natexlab#1{#1}\fi
\expandafter\ifx\csname bibnamefont\endcsname\relax
  \def\bibnamefont#1{#1}\fi
\expandafter\ifx\csname bibfnamefont\endcsname\relax
  \def\bibfnamefont#1{#1}\fi
\expandafter\ifx\csname citenamefont\endcsname\relax
  \def\citenamefont#1{#1}\fi
\expandafter\ifx\csname url\endcsname\relax
  \def\url#1{\texttt{#1}}\fi
\expandafter\ifx\csname urlprefix\endcsname\relax\def\urlprefix{URL }\fi
\providecommand{\bibinfo}[2]{#2}
\providecommand{\eprint}[2][]{\url{#2}}

\bibitem[{\citenamefont{Mention et~al.}(2011)\citenamefont{Mention, Fechner,
  Lasserre, Mueller, Lhuillier, Cribier, and Letourneau}}]{Mention:2011rk}
\bibinfo{author}{\bibfnamefont{G.}~\bibnamefont{Mention}},
  \bibinfo{author}{\bibfnamefont{M.}~\bibnamefont{Fechner}},
  \bibinfo{author}{\bibfnamefont{{\relax Th}.}~\bibnamefont{Lasserre}},
  \bibinfo{author}{\bibfnamefont{{\relax Th}.~A.} \bibnamefont{Mueller}},
  \bibinfo{author}{\bibfnamefont{D.}~\bibnamefont{Lhuillier}},
  \bibinfo{author}{\bibfnamefont{M.}~\bibnamefont{Cribier}}, \bibnamefont{and}
  \bibinfo{author}{\bibfnamefont{A.}~\bibnamefont{Letourneau}}, ``{The Reactor
  Antineutrino Anomaly},'' \bibinfo{journal}{Phys. Rev.}
  \textbf{\bibinfo{volume}{D83}}, \bibinfo{pages}{073006}
  (\bibinfo{year}{2011}), \eprint{1101.2755}.

\bibitem[{\citenamefont{Hampel et~al.}(1998)}]{Hampel:1997fc}
\bibinfo{author}{\bibfnamefont{W.}~\bibnamefont{Hampel}} \bibnamefont{et~al.}
  (\bibinfo{collaboration}{GALLEX}), ``{Final results of the $^{51}$Cr neutrino
  source experiments in GALLEX},'' \bibinfo{journal}{Phys. Lett.}
  \textbf{\bibinfo{volume}{B420}}, \bibinfo{pages}{114} (\bibinfo{year}{1998}).

\bibitem[{\citenamefont{Abdurashitov et~al.}(1999)}]{Abdurashitov:1998ne}
\bibinfo{author}{\bibfnamefont{J.~N.} \bibnamefont{Abdurashitov}}
  \bibnamefont{et~al.} (\bibinfo{collaboration}{SAGE}), ``{Measurement of the
  response of the Russian-American gallium experiment to neutrinos from a
  $^{51}$Cr source},'' \bibinfo{journal}{Phys. Rev.}
  \textbf{\bibinfo{volume}{C59}}, \bibinfo{pages}{2246} (\bibinfo{year}{1999}),
  \eprint{hep-ph/9803418}.

\bibitem[{\citenamefont{Aguilar-Arevalo et~al.}(2001)}]{Aguilar:2001ty}
\bibinfo{author}{\bibfnamefont{A.}~\bibnamefont{Aguilar-Arevalo}}
  \bibnamefont{et~al.} (\bibinfo{collaboration}{LSND}), ``{Evidence for
  neutrino oscillations from the observation of $\overline{\nu}_e$ appearance
  in a $\overline{\nu}_\mu$ beam},'' \bibinfo{journal}{Phys. Rev.}
  \textbf{\bibinfo{volume}{D64}}, \bibinfo{pages}{112007}
  (\bibinfo{year}{2001}), \eprint{hep-ex/0104049}.

\bibitem[{\citenamefont{Aguilar-Arevalo
  et~al.}(2013)}]{Aguilar-Arevalo:2013pmq}
\bibinfo{author}{\bibfnamefont{A.~A.} \bibnamefont{Aguilar-Arevalo}}
  \bibnamefont{et~al.} (\bibinfo{collaboration}{MiniBooNE}), ``{Improved Search
  for $\overline \nu_\mu \rightarrow \overline \nu_e$ Oscillations in the
  MiniBooNE Experiment},'' \bibinfo{journal}{Phys. Rev. Lett.}
  \textbf{\bibinfo{volume}{110}}, \bibinfo{pages}{161801}
  (\bibinfo{year}{2013}), \eprint{1303.2588}.

\bibitem[{\citenamefont{Dentler et~al.}(2018)\citenamefont{Dentler,
  Hern\'{a}ndez-Cabezudo, Kopp, Machado, Maltoni, Martinez-Soler, and
  Schwetz}}]{Dentler:2018sju}
\bibinfo{author}{\bibfnamefont{M.}~\bibnamefont{Dentler}},
  \bibinfo{author}{\bibfnamefont{A.}~\bibnamefont{Hern\'{a}ndez-Cabezudo}},
  \bibinfo{author}{\bibfnamefont{J.}~\bibnamefont{Kopp}},
  \bibinfo{author}{\bibfnamefont{P.~A.~N.} \bibnamefont{Machado}},
  \bibinfo{author}{\bibfnamefont{M.}~\bibnamefont{Maltoni}},
  \bibinfo{author}{\bibfnamefont{I.}~\bibnamefont{Martinez-Soler}},
  \bibnamefont{and} \bibinfo{author}{\bibfnamefont{T.}~\bibnamefont{Schwetz}},
  ``{Updated Global Analysis of Neutrino Oscillations in the Presence of
  eV-Scale Sterile Neutrinos},'' \bibinfo{journal}{JHEP}
  \textbf{\bibinfo{volume}{08}}, \bibinfo{pages}{010} (\bibinfo{year}{2018}),
  \eprint{1803.10661}.

\bibitem[{\citenamefont{Diaz et~al.}(2019)\citenamefont{Diaz, Arg{\"u}elles,
  Collin, Conrad, and Shaevitz}}]{Diaz:2019fwt}
\bibinfo{author}{\bibfnamefont{A.}~\bibnamefont{Diaz}},
  \bibinfo{author}{\bibfnamefont{C.~A.} \bibnamefont{Arg{\"u}elles}},
  \bibinfo{author}{\bibfnamefont{G.~H.} \bibnamefont{Collin}},
  \bibinfo{author}{\bibfnamefont{J.~M.} \bibnamefont{Conrad}},
  \bibnamefont{and} \bibinfo{author}{\bibfnamefont{M.~H.}
  \bibnamefont{Shaevitz}}, ``{Where Are We With Light Sterile Neutrinos?},''
  (\bibinfo{year}{2019}), \eprint{1906.00045}.

\bibitem[{\citenamefont{B{\"o}ser et~al.}(2019)\citenamefont{B{\"o}ser, Buck,
  Giunti, Lesgourgues, Ludhova, Mertens, Schukraft, and Wurm}}]{Boser:2019rta}
\bibinfo{author}{\bibfnamefont{S.}~\bibnamefont{B{\"o}ser}},
  \bibinfo{author}{\bibfnamefont{C.}~\bibnamefont{Buck}},
  \bibinfo{author}{\bibfnamefont{C.}~\bibnamefont{Giunti}},
  \bibinfo{author}{\bibfnamefont{J.}~\bibnamefont{Lesgourgues}},
  \bibinfo{author}{\bibfnamefont{L.}~\bibnamefont{Ludhova}},
  \bibinfo{author}{\bibfnamefont{S.}~\bibnamefont{Mertens}},
  \bibinfo{author}{\bibfnamefont{A.}~\bibnamefont{Schukraft}},
  \bibnamefont{and} \bibinfo{author}{\bibfnamefont{M.}~\bibnamefont{Wurm}},
  ``{Status of Light Sterile Neutrino Searches},''  (\bibinfo{year}{2019}),
  \eprint{1906.01739}.

\bibitem[{\citenamefont{Merle et~al.}(2014)\citenamefont{Merle, Morisi, and
  Winter}}]{Merle:2014eja}
\bibinfo{author}{\bibfnamefont{A.}~\bibnamefont{Merle}},
  \bibinfo{author}{\bibfnamefont{S.}~\bibnamefont{Morisi}}, \bibnamefont{and}
  \bibinfo{author}{\bibfnamefont{W.}~\bibnamefont{Winter}}, ``{Common origin of
  reactor and sterile neutrino mixing},'' \bibinfo{journal}{JHEP}
  \textbf{\bibinfo{volume}{07}}, \bibinfo{pages}{039} (\bibinfo{year}{2014}),
  \eprint{1402.6332}.

\bibitem[{\citenamefont{Bakhti et~al.}(2015)\citenamefont{Bakhti, Farzan, and
  Schwetz}}]{Bakhti:2015dca}
\bibinfo{author}{\bibfnamefont{P.}~\bibnamefont{Bakhti}},
  \bibinfo{author}{\bibfnamefont{Y.}~\bibnamefont{Farzan}}, \bibnamefont{and}
  \bibinfo{author}{\bibfnamefont{T.}~\bibnamefont{Schwetz}}, ``{Revisiting the
  quantum decoherence scenario as an explanation for the LSND anomaly},''
  \bibinfo{journal}{JHEP} \textbf{\bibinfo{volume}{05}}, \bibinfo{pages}{007}
  (\bibinfo{year}{2015}), \eprint{1503.05374}.

\bibitem[{\citenamefont{Babu et~al.}(2016)\citenamefont{Babu, McKay, Mocioiu,
  and Pakvasa}}]{Babu:2016fdt}
\bibinfo{author}{\bibfnamefont{K.~S.} \bibnamefont{Babu}},
  \bibinfo{author}{\bibfnamefont{D.~W.} \bibnamefont{McKay}},
  \bibinfo{author}{\bibfnamefont{I.}~\bibnamefont{Mocioiu}}, \bibnamefont{and}
  \bibinfo{author}{\bibfnamefont{S.}~\bibnamefont{Pakvasa}}, ``{Light sterile
  neutrinos, lepton number violating interactions, and the LSND neutrino
  anomaly},'' \bibinfo{journal}{Phys. Rev.} \textbf{\bibinfo{volume}{D93}},
  \bibinfo{pages}{113019} (\bibinfo{year}{2016}), \eprint{1605.03625}.

\bibitem[{\citenamefont{Carena et~al.}(2017)\citenamefont{Carena, Li, Machado,
  Machado, and Wagner}}]{Carena:2017qhd}
\bibinfo{author}{\bibfnamefont{M.}~\bibnamefont{Carena}},
  \bibinfo{author}{\bibfnamefont{Y.-Y.} \bibnamefont{Li}},
  \bibinfo{author}{\bibfnamefont{C.~S.} \bibnamefont{Machado}},
  \bibinfo{author}{\bibfnamefont{P.~A.~N.} \bibnamefont{Machado}},
  \bibnamefont{and} \bibinfo{author}{\bibfnamefont{C.~E.~M.}
  \bibnamefont{Wagner}}, ``{Neutrinos in Large Extra Dimensions and
  Short-Baseline $\nu_e$ Appearance},'' \bibinfo{journal}{Phys. Rev.}
  \textbf{\bibinfo{volume}{D96}}, \bibinfo{pages}{095014}
  (\bibinfo{year}{2017}), \eprint{1708.09548}.

\bibitem[{\citenamefont{Magill et~al.}(2018)\citenamefont{Magill, Plestid,
  Pospelov, and Tsai}}]{Magill:2018jla}
\bibinfo{author}{\bibfnamefont{G.}~\bibnamefont{Magill}},
  \bibinfo{author}{\bibfnamefont{R.}~\bibnamefont{Plestid}},
  \bibinfo{author}{\bibfnamefont{M.}~\bibnamefont{Pospelov}}, \bibnamefont{and}
  \bibinfo{author}{\bibfnamefont{Y.-D.} \bibnamefont{Tsai}}, ``{Dipole Portal
  to Heavy Neutral Leptons},'' \bibinfo{journal}{Phys. Rev.}
  \textbf{\bibinfo{volume}{D98}}, \bibinfo{pages}{115015}
  (\bibinfo{year}{2018}), \eprint{1803.03262}.

\bibitem[{\citenamefont{D{\"o}ring et~al.}(2018)\citenamefont{D{\"o}ring,
  P{\"a}s, Sicking, and Weiler}}]{Doring:2018cob}
\bibinfo{author}{\bibfnamefont{D.}~\bibnamefont{D{\"o}ring}},
  \bibinfo{author}{\bibfnamefont{H.}~\bibnamefont{P{\"a}s}},
  \bibinfo{author}{\bibfnamefont{P.}~\bibnamefont{Sicking}}, \bibnamefont{and}
  \bibinfo{author}{\bibfnamefont{T.~J.} \bibnamefont{Weiler}}, ``{Sterile
  Neutrinos with Altered Dispersion Relations as an Explanation for the
  MiniBooNE, LSND, Gallium and Reactor Anomalies},''  (\bibinfo{year}{2018}),
  \eprint{1808.07460}.

\bibitem[{\citenamefont{Liao et~al.}(2019)\citenamefont{Liao, Marfatia, and
  Whisnant}}]{Liao:2018mbg}
\bibinfo{author}{\bibfnamefont{J.}~\bibnamefont{Liao}},
  \bibinfo{author}{\bibfnamefont{D.}~\bibnamefont{Marfatia}}, \bibnamefont{and}
  \bibinfo{author}{\bibfnamefont{K.}~\bibnamefont{Whisnant}}, ``{MiniBooNE,
  MINOS+ and IceCube data imply a baroque neutrino sector},''
  \bibinfo{journal}{Phys. Rev.} \textbf{\bibinfo{volume}{D99}},
  \bibinfo{pages}{015016} (\bibinfo{year}{2019}), \eprint{1810.01000}.

\bibitem[{\citenamefont{Denton et~al.}(2019)\citenamefont{Denton, Farzan, and
  Shoemaker}}]{Denton:2018dqq}
\bibinfo{author}{\bibfnamefont{P.~B.} \bibnamefont{Denton}},
  \bibinfo{author}{\bibfnamefont{Y.}~\bibnamefont{Farzan}}, \bibnamefont{and}
  \bibinfo{author}{\bibfnamefont{I.~M.} \bibnamefont{Shoemaker}}, ``{Activating
  the fourth neutrino of the 3+1 scheme},'' \bibinfo{journal}{Phys. Rev.}
  \textbf{\bibinfo{volume}{D99}}, \bibinfo{pages}{035003}
  (\bibinfo{year}{2019}), \eprint{1811.01310}.

\bibitem[{\citenamefont{Kostensalo et~al.}(2019)\citenamefont{Kostensalo,
  Suhonen, Giunti, and Srivastava}}]{Kostensalo:2019vmv}
\bibinfo{author}{\bibfnamefont{J.}~\bibnamefont{Kostensalo}},
  \bibinfo{author}{\bibfnamefont{J.}~\bibnamefont{Suhonen}},
  \bibinfo{author}{\bibfnamefont{C.}~\bibnamefont{Giunti}}, \bibnamefont{and}
  \bibinfo{author}{\bibfnamefont{P.~C.} \bibnamefont{Srivastava}}, ``{The
  gallium anomaly revisited},'' \bibinfo{journal}{Phys. Lett.}
  \textbf{\bibinfo{volume}{B795}}, \bibinfo{pages}{542} (\bibinfo{year}{2019}),
  \eprint{1906.10980}.

\bibitem[{\citenamefont{Antonello et~al.}(2015)}]{Antonello:2015lea}
\bibinfo{author}{\bibfnamefont{M.}~\bibnamefont{Antonello}}
  \bibnamefont{et~al.} (\bibinfo{collaboration}{MicroBooNE, LAr1-ND,
  ICARUS-WA104}), ``{A Proposal for a Three Detector Short-Baseline Neutrino
  Oscillation Program in the Fermilab Booster Neutrino Beam},''
  (\bibinfo{year}{2015}), \eprint{1503.01520}.

\bibitem[{\citenamefont{Machado et~al.}(2019)\citenamefont{Machado, Palamara,
  and Schmitz}}]{Machado:2019oxb}
\bibinfo{author}{\bibfnamefont{P.~A.} \bibnamefont{Machado}},
  \bibinfo{author}{\bibfnamefont{O.}~\bibnamefont{Palamara}}, \bibnamefont{and}
  \bibinfo{author}{\bibfnamefont{D.~W.} \bibnamefont{Schmitz}}, ``{The
  Short-Baseline Neutrino Program at Fermilab},'' \bibinfo{journal}{Ann. Rev.
  Nucl. Part. Sci.} \textbf{\bibinfo{volume}{69}} (\bibinfo{year}{2019}),
  \eprint{1903.04608}.

\bibitem[{\citenamefont{Huber et~al.}(2005)\citenamefont{Huber, Lindner, and
  Winter}}]{Huber:2004ka}
\bibinfo{author}{\bibfnamefont{P.}~\bibnamefont{Huber}},
  \bibinfo{author}{\bibfnamefont{M.}~\bibnamefont{Lindner}}, \bibnamefont{and}
  \bibinfo{author}{\bibfnamefont{W.}~\bibnamefont{Winter}}, ``{Simulation of
  long-baseline neutrino oscillation experiments with GLoBES (General Long
  Baseline Experiment Simulator)},'' \bibinfo{journal}{Comput. Phys. Commun.}
  \textbf{\bibinfo{volume}{167}}, \bibinfo{pages}{195} (\bibinfo{year}{2005}),
  \eprint{hep-ph/0407333}.

\bibitem[{\citenamefont{Huber et~al.}(2007)\citenamefont{Huber, Kopp, Lindner,
  Rolinec, and Winter}}]{Huber:2007ji}
\bibinfo{author}{\bibfnamefont{P.}~\bibnamefont{Huber}},
  \bibinfo{author}{\bibfnamefont{J.}~\bibnamefont{Kopp}},
  \bibinfo{author}{\bibfnamefont{M.}~\bibnamefont{Lindner}},
  \bibinfo{author}{\bibfnamefont{M.}~\bibnamefont{Rolinec}}, \bibnamefont{and}
  \bibinfo{author}{\bibfnamefont{W.}~\bibnamefont{Winter}}, ``{New features in
  the simulation of neutrino oscillation experiments with GLoBES 3.0: General
  Long Baseline Experiment Simulator},'' \bibinfo{journal}{Comput. Phys.
  Commun.} \textbf{\bibinfo{volume}{177}}, \bibinfo{pages}{432}
  (\bibinfo{year}{2007}), \eprint{hep-ph/0701187}.

\bibitem[{\citenamefont{Berryman and Huber}()}]{upcoming}
\bibinfo{author}{\bibfnamefont{J.~M.} \bibnamefont{Berryman}} \bibnamefont{and}
  \bibinfo{author}{\bibfnamefont{P.}~\bibnamefont{Huber}}, \bibinfo{note}{in
  Preparation}.

\bibitem[{\citenamefont{Mueller et~al.}(2011)}]{Mueller:2011nm}
\bibinfo{author}{\bibfnamefont{{\relax Th}.~A.} \bibnamefont{Mueller}}
  \bibnamefont{et~al.}, ``{Improved Predictions of Reactor Antineutrino
  Spectra},'' \bibinfo{journal}{Phys. Rev.} \textbf{\bibinfo{volume}{C83}},
  \bibinfo{pages}{054615} (\bibinfo{year}{2011}), \eprint{1101.2663}.

\bibitem[{\citenamefont{Huber}(2011)}]{Huber:2011wv}
\bibinfo{author}{\bibfnamefont{P.}~\bibnamefont{Huber}}, ``{On the
  determination of anti-neutrino spectra from nuclear reactors},''
  \bibinfo{journal}{Phys. Rev.} \textbf{\bibinfo{volume}{C84}},
  \bibinfo{pages}{024617} (\bibinfo{year}{2011}), \bibinfo{note}{[Erratum:
  Phys. Rev. C85, 029901 (2012)]}, \eprint{1106.0687}.

\bibitem[{\citenamefont{Von~Feilitzsch
  et~al.}(1982)\citenamefont{Von~Feilitzsch, Hahn, and
  Schreckenbach}}]{VonFeilitzsch:1982jw}
\bibinfo{author}{\bibfnamefont{F.}~\bibnamefont{Von~Feilitzsch}},
  \bibinfo{author}{\bibfnamefont{A.~A.} \bibnamefont{Hahn}}, \bibnamefont{and}
  \bibinfo{author}{\bibfnamefont{K.}~\bibnamefont{Schreckenbach}},
  ``{Experimental beta-spectra from $^{239}$Pu and $^{235}$U thermal neutron
  fission products and their correlated antineutrino spectra},''
  \bibinfo{journal}{Phys. Lett.} \textbf{\bibinfo{volume}{118B}},
  \bibinfo{pages}{162} (\bibinfo{year}{1982}).

\bibitem[{\citenamefont{Schreckenbach et~al.}(1985)\citenamefont{Schreckenbach,
  Colvin, Gelletly, and Von~Feilitzsch}}]{Schreckenbach:1985ep}
\bibinfo{author}{\bibfnamefont{K.}~\bibnamefont{Schreckenbach}},
  \bibinfo{author}{\bibfnamefont{G.}~\bibnamefont{Colvin}},
  \bibinfo{author}{\bibfnamefont{W.}~\bibnamefont{Gelletly}}, \bibnamefont{and}
  \bibinfo{author}{\bibfnamefont{F.}~\bibnamefont{Von~Feilitzsch}},
  ``{Determination of the antineutrino spectrum from $^{235}$U thermal neutron
  fission products up to 9.5 MeV},'' \bibinfo{journal}{Phys. Lett.}
  \textbf{\bibinfo{volume}{160B}}, \bibinfo{pages}{325} (\bibinfo{year}{1985}).

\bibitem[{\citenamefont{Hahn et~al.}(1989)\citenamefont{Hahn, Schreckenbach,
  Colvin, Krusche, Gelletly, and Von~Feilitzsch}}]{Hahn:1989zr}
\bibinfo{author}{\bibfnamefont{A.~A.} \bibnamefont{Hahn}},
  \bibinfo{author}{\bibfnamefont{K.}~\bibnamefont{Schreckenbach}},
  \bibinfo{author}{\bibfnamefont{G.}~\bibnamefont{Colvin}},
  \bibinfo{author}{\bibfnamefont{B.}~\bibnamefont{Krusche}},
  \bibinfo{author}{\bibfnamefont{W.}~\bibnamefont{Gelletly}}, \bibnamefont{and}
  \bibinfo{author}{\bibfnamefont{F.}~\bibnamefont{Von~Feilitzsch}},
  ``{Antineutrino spectra from $^{241}$Pu and $^{239}$Pu thermal neutron
  fission products},'' \bibinfo{journal}{Phys. Lett.}
  \textbf{\bibinfo{volume}{B218}}, \bibinfo{pages}{365} (\bibinfo{year}{1989}).

\bibitem[{\citenamefont{Estienne et~al.}(2019)}]{Estienne:2019ujo}
\bibinfo{author}{\bibfnamefont{M.}~\bibnamefont{Estienne}}
  \bibnamefont{et~al.}, ``{Updated Summation Model: An Improved Agreement with
  the Daya Bay Antineutrino Fluxes},'' \bibinfo{journal}{Phys. Rev. Lett.}
  \textbf{\bibinfo{volume}{123}}, \bibinfo{pages}{022502}
  (\bibinfo{year}{2019}), \eprint{1904.09358}.

\bibitem[{\citenamefont{Gariazzo et~al.}(2018)\citenamefont{Gariazzo, Giunti,
  Laveder, and Li}}]{Gariazzo:2018mwd}
\bibinfo{author}{\bibfnamefont{S.}~\bibnamefont{Gariazzo}},
  \bibinfo{author}{\bibfnamefont{C.}~\bibnamefont{Giunti}},
  \bibinfo{author}{\bibfnamefont{M.}~\bibnamefont{Laveder}}, \bibnamefont{and}
  \bibinfo{author}{\bibfnamefont{Y.~F.} \bibnamefont{Li}}, ``{Model-independent
  $\bar\nu_{e}$ short-baseline oscillations from reactor spectral ratios},''
  \bibinfo{journal}{Phys. Lett.} \textbf{\bibinfo{volume}{B782}},
  \bibinfo{pages}{13} (\bibinfo{year}{2018}), \eprint{1801.06467}.

\bibitem[{\citenamefont{Adey et~al.}(2019)}]{Adey:2019ywk}
\bibinfo{author}{\bibfnamefont{D.}~\bibnamefont{Adey}} \bibnamefont{et~al.}
  (\bibinfo{collaboration}{Daya Bay}), ``{Measurement of Individual
  Antineutrino Spectra from $\mathbf{^{235}U}$ and $\mathbf{^{239}Pu}$ at Daya
  Bay},''  (\bibinfo{year}{2019}), \eprint{1904.07812}.

\bibitem[{\citenamefont{Hayen et~al.}(2019)\citenamefont{Hayen, Kostensalo,
  Severijns, and Suhonen}}]{Hayen:2019eop}
\bibinfo{author}{\bibfnamefont{L.}~\bibnamefont{Hayen}},
  \bibinfo{author}{\bibfnamefont{J.}~\bibnamefont{Kostensalo}},
  \bibinfo{author}{\bibfnamefont{N.}~\bibnamefont{Severijns}},
  \bibnamefont{and} \bibinfo{author}{\bibfnamefont{J.}~\bibnamefont{Suhonen}},
  ``{First-forbidden transitions in the reactor anomaly},''
  (\bibinfo{year}{2019}), \eprint{1908.08302}.

\bibitem[{\citenamefont{Ko et~al.}(2017)}]{Ko:2016owz}
\bibinfo{author}{\bibfnamefont{Y.}~\bibnamefont{Ko}} \bibnamefont{et~al.}
  (\bibinfo{collaboration}{NEOS}), ``{Sterile Neutrino Search at the NEOS
  Experiment},'' \bibinfo{journal}{Phys. Rev. Lett.}
  \textbf{\bibinfo{volume}{118}}, \bibinfo{pages}{121802}
  (\bibinfo{year}{2017}), \eprint{1610.05134}.

\bibitem[{\citenamefont{Bak et~al.}(2018{\natexlab{a}})}]{Bak:2018ydk}
\bibinfo{author}{\bibfnamefont{G.}~\bibnamefont{Bak}} \bibnamefont{et~al.}
  (\bibinfo{collaboration}{RENO}), ``{Measurement of Reactor Antineutrino
  Oscillation Amplitude and Frequency at RENO},'' \bibinfo{journal}{Phys. Rev.
  Lett.} \textbf{\bibinfo{volume}{121}}, \bibinfo{pages}{201801}
  (\bibinfo{year}{2018}{\natexlab{a}}), \eprint{1806.00248}.

\bibitem[{\citenamefont{De~Kerret et~al.}(2019)}]{DoubleChooz:2019qbj}
\bibinfo{author}{\bibfnamefont{H.}~\bibnamefont{De~Kerret}}
  \bibnamefont{et~al.} (\bibinfo{collaboration}{Double Chooz}), ``{First Double
  Chooz $\mathbf{\theta_{13}}$ Measurement via Total Neutron Capture
  Detection},''  (\bibinfo{year}{2019}), \eprint{1901.09445}.

\bibitem[{\citenamefont{Declais et~al.}(1994)}]{Declais:1994ma}
\bibinfo{author}{\bibfnamefont{Y.}~\bibnamefont{Declais}} \bibnamefont{et~al.},
  ``{Study of reactor anti-neutrino interaction with proton at Bugey nuclear
  power plant},'' \bibinfo{journal}{Phys. Lett.}
  \textbf{\bibinfo{volume}{B338}}, \bibinfo{pages}{383} (\bibinfo{year}{1994}).

\bibitem[{\citenamefont{Declais et~al.}(1995)}]{Declais:1994su}
\bibinfo{author}{\bibfnamefont{Y.}~\bibnamefont{Declais}} \bibnamefont{et~al.},
  ``{Search for neutrino oscillations at 15-meters, 40-meters, and 95-meters
  from a nuclear power reactor at Bugey},'' \bibinfo{journal}{Nucl. Phys.}
  \textbf{\bibinfo{volume}{B434}}, \bibinfo{pages}{503} (\bibinfo{year}{1995}).

\bibitem[{\citenamefont{Zacek et~al.}(1986)}]{Zacek:1986cu}
\bibinfo{author}{\bibfnamefont{G.}~\bibnamefont{Zacek}} \bibnamefont{et~al.}
  (\bibinfo{collaboration}{CALTECH-SIN-TUM}), ``{Neutrino Oscillation
  Experiments at the G\"{o}sgen Nuclear Power Reactor},''
  \bibinfo{journal}{Phys. Rev.} \textbf{\bibinfo{volume}{D34}},
  \bibinfo{pages}{2621} (\bibinfo{year}{1986}).

\bibitem[{\citenamefont{Kwon et~al.}(1981)\citenamefont{Kwon, Boehm, Hahn,
  Henrikson, Vuilleumier, Cavaignac, Koang, Vignon, Von~Feilitzsch, and
  Mossbauer}}]{Kwon:1981ua}
\bibinfo{author}{\bibfnamefont{H.}~\bibnamefont{Kwon}},
  \bibinfo{author}{\bibfnamefont{F.}~\bibnamefont{Boehm}},
  \bibinfo{author}{\bibfnamefont{A.~A.} \bibnamefont{Hahn}},
  \bibinfo{author}{\bibfnamefont{H.~E.} \bibnamefont{Henrikson}},
  \bibinfo{author}{\bibfnamefont{J.~L.} \bibnamefont{Vuilleumier}},
  \bibinfo{author}{\bibfnamefont{J.~F.} \bibnamefont{Cavaignac}},
  \bibinfo{author}{\bibfnamefont{D.~H.} \bibnamefont{Koang}},
  \bibinfo{author}{\bibfnamefont{B.}~\bibnamefont{Vignon}},
  \bibinfo{author}{\bibfnamefont{F.}~\bibnamefont{Von~Feilitzsch}},
  \bibnamefont{and} \bibinfo{author}{\bibfnamefont{R.~L.}
  \bibnamefont{Mossbauer}}, ``{Search for Neutrino Oscillations at a Fission
  Reactor},'' \bibinfo{journal}{Phys. Rev.} \textbf{\bibinfo{volume}{D24}},
  \bibinfo{pages}{1097} (\bibinfo{year}{1981}).

\bibitem[{\citenamefont{Hoummada et~al.}(1995)\citenamefont{Hoummada,
  Lazrak~Mikou, Avenier, Bagieu, Cavaignac, and Holm~Koan}}]{ILLupdate}
\bibinfo{author}{\bibfnamefont{A.}~\bibnamefont{Hoummada}},
  \bibinfo{author}{\bibfnamefont{S.}~\bibnamefont{Lazrak~Mikou}},
  \bibinfo{author}{\bibfnamefont{M.}~\bibnamefont{Avenier}},
  \bibinfo{author}{\bibfnamefont{G.}~\bibnamefont{Bagieu}},
  \bibinfo{author}{\bibfnamefont{J.~F.} \bibnamefont{Cavaignac}},
  \bibnamefont{and} \bibinfo{author}{\bibfnamefont{{\relax
  Dy}.}~\bibnamefont{Holm~Koan}}, ``{Neutrino oscillations I.L.L. experiment
  reanalysis},'' \bibinfo{journal}{Appl. Radiat. Isot.}
  \textbf{\bibinfo{volume}{46}}, \bibinfo{pages}{449} (\bibinfo{year}{1995}).

\bibitem[{\citenamefont{Vidyakin et~al.}(1987)\citenamefont{Vidyakin, Vyrodov,
  Gurevich, Kozlov, Martemyanov, Sukhotin, Tarasenkov, and
  Khakimov}}]{Vidyakin:1987ue}
\bibinfo{author}{\bibfnamefont{G.~S.} \bibnamefont{Vidyakin}},
  \bibinfo{author}{\bibfnamefont{V.~N.} \bibnamefont{Vyrodov}},
  \bibinfo{author}{\bibfnamefont{I.~I.} \bibnamefont{Gurevich}},
  \bibinfo{author}{\bibfnamefont{{\relax Yu}.~V.} \bibnamefont{Kozlov}},
  \bibinfo{author}{\bibfnamefont{V.~P.} \bibnamefont{Martemyanov}},
  \bibinfo{author}{\bibfnamefont{S.~V.} \bibnamefont{Sukhotin}},
  \bibinfo{author}{\bibfnamefont{V.~G.} \bibnamefont{Tarasenkov}},
  \bibnamefont{and} \bibinfo{author}{\bibfnamefont{S.~K.}
  \bibnamefont{Khakimov}}, ``{Detection of Anti-neutrinos in the Flux From Two
  Reactors},'' \bibinfo{journal}{Sov. Phys. JETP}
  \textbf{\bibinfo{volume}{66}}, \bibinfo{pages}{243} (\bibinfo{year}{1987}),
  \bibinfo{note}{[Zh. Eksp. Teor. Fiz. 93, 424 (1987)]}.

\bibitem[{\citenamefont{Vidyakin et~al.}(1994)}]{Vidyakin:1994ut}
\bibinfo{author}{\bibfnamefont{G.~S.} \bibnamefont{Vidyakin}}
  \bibnamefont{et~al.}, ``{Limitations on the characteristics of neutrino
  oscillations},'' \bibinfo{journal}{JETP Lett.} \textbf{\bibinfo{volume}{59}},
  \bibinfo{pages}{390} (\bibinfo{year}{1994}), \bibinfo{note}{[Pisma Zh. Eksp.
  Teor. Fiz. 59, 364 (1994)]}.

\bibitem[{\citenamefont{Kozlov et~al.}(2000)\citenamefont{Kozlov, Khalturtsev,
  Machulin, Martemyanov, Martemyanov, Sukhotin, Tarasenkov, Turbin, and
  Vyrodov}}]{Kozlov:1999ct}
\bibinfo{author}{\bibfnamefont{{\relax Yu}.~V.} \bibnamefont{Kozlov}},
  \bibinfo{author}{\bibfnamefont{S.~V.} \bibnamefont{Khalturtsev}},
  \bibinfo{author}{\bibfnamefont{I.~N.} \bibnamefont{Machulin}},
  \bibinfo{author}{\bibfnamefont{A.~V.} \bibnamefont{Martemyanov}},
  \bibinfo{author}{\bibfnamefont{V.~P.} \bibnamefont{Martemyanov}},
  \bibinfo{author}{\bibfnamefont{S.~V.} \bibnamefont{Sukhotin}},
  \bibinfo{author}{\bibfnamefont{V.~G.} \bibnamefont{Tarasenkov}},
  \bibinfo{author}{\bibfnamefont{E.~V.} \bibnamefont{Turbin}},
  \bibnamefont{and} \bibinfo{author}{\bibfnamefont{V.~N.}
  \bibnamefont{Vyrodov}}, ``{Antineutrino-Deuteron Experiment at the
  Krasnoyarsk Reactor},'' \bibinfo{journal}{Phys. Atom. Nucl.}
  \textbf{\bibinfo{volume}{63}}, \bibinfo{pages}{1016} (\bibinfo{year}{2000}),
  \bibinfo{note}{[Yad. Fiz. 63, 1091 (2000)]}, \eprint{hep-ex/9912047}.

\bibitem[{\citenamefont{Boireau et~al.}(2016)}]{Boireau:2015dda}
\bibinfo{author}{\bibfnamefont{G.}~\bibnamefont{Boireau}} \bibnamefont{et~al.}
  (\bibinfo{collaboration}{NUCIFER}), ``{Online Monitoring of the Osiris
  Reactor with the Nucifer Neutrino Detector},'' \bibinfo{journal}{Phys. Rev.}
  \textbf{\bibinfo{volume}{D93}}, \bibinfo{pages}{112006}
  (\bibinfo{year}{2016}), \eprint{1509.05610}.

\bibitem[{\citenamefont{Greenwood et~al.}(1996)}]{Greenwood:1996pb}
\bibinfo{author}{\bibfnamefont{Z.~D.} \bibnamefont{Greenwood}}
  \bibnamefont{et~al.}, ``{Results of a two position reactor neutrino
  oscillation experiment},'' \bibinfo{journal}{Phys. Rev.}
  \textbf{\bibinfo{volume}{D53}}, \bibinfo{pages}{6054} (\bibinfo{year}{1996}).

\bibitem[{\citenamefont{Afonin et~al.}(1988)\citenamefont{Afonin, Ketov,
  Kopeikin, Mikaelyan, Skorokhvatov, and Tolokonnikov}}]{Afonin:1988gx}
\bibinfo{author}{\bibfnamefont{A.~I.} \bibnamefont{Afonin}},
  \bibinfo{author}{\bibfnamefont{S.~N.} \bibnamefont{Ketov}},
  \bibinfo{author}{\bibfnamefont{V.~I.} \bibnamefont{Kopeikin}},
  \bibinfo{author}{\bibfnamefont{L.~A.} \bibnamefont{Mikaelyan}},
  \bibinfo{author}{\bibfnamefont{M.~D.} \bibnamefont{Skorokhvatov}},
  \bibnamefont{and} \bibinfo{author}{\bibfnamefont{S.~V.}
  \bibnamefont{Tolokonnikov}}, ``{A Study of the Reaction $\bar{\nu}_e + P \to
  e^+ + N$ on a Nuclear Reactor},'' \bibinfo{journal}{Sov. Phys. JETP}
  \textbf{\bibinfo{volume}{67}}, \bibinfo{pages}{213} (\bibinfo{year}{1988}),
  \bibinfo{note}{[Zh. Eksp. Teor. Fiz. 94 N2,1 (1988)]}.

\bibitem[{\citenamefont{Kuvshinnikov et~al.}(1991)\citenamefont{Kuvshinnikov,
  Mikaelyan, Nikolaev, Skorokhvatov, and Etenko}}]{Kuvshinnikov:1990ry}
\bibinfo{author}{\bibfnamefont{A.~A.} \bibnamefont{Kuvshinnikov}},
  \bibinfo{author}{\bibfnamefont{L.~A.} \bibnamefont{Mikaelyan}},
  \bibinfo{author}{\bibfnamefont{S.~V.} \bibnamefont{Nikolaev}},
  \bibinfo{author}{\bibfnamefont{M.~D.} \bibnamefont{Skorokhvatov}},
  \bibnamefont{and} \bibinfo{author}{\bibfnamefont{A.~V.}
  \bibnamefont{Etenko}}, ``{Measuring the $\overline{\nu}_e + p \to n + e^+$
  cross-section and beta decay axial constant in a new experiment at Rovno NPP
  reactor. (In Russian)},'' \bibinfo{journal}{JETP Lett.}
  \textbf{\bibinfo{volume}{54}}, \bibinfo{pages}{253} (\bibinfo{year}{1991}),
  \bibinfo{note}{[Sov. J. Nucl. Phys. 52, 300 (1990)]}.

\bibitem[{\citenamefont{Apollonio et~al.}(2003)}]{Apollonio:2002gd}
\bibinfo{author}{\bibfnamefont{M.}~\bibnamefont{Apollonio}}
  \bibnamefont{et~al.} (\bibinfo{collaboration}{CHOOZ}), ``{Search for neutrino
  oscillations on a long baseline at the CHOOZ nuclear power station},''
  \bibinfo{journal}{Eur. Phys. J.} \textbf{\bibinfo{volume}{C27}},
  \bibinfo{pages}{331} (\bibinfo{year}{2003}), \eprint{hep-ex/0301017}.

\bibitem[{\citenamefont{Boehm et~al.}(2001)}]{Boehm:2001ik}
\bibinfo{author}{\bibfnamefont{F.}~\bibnamefont{Boehm}} \bibnamefont{et~al.},
  ``{Final results from the Palo Verde neutrino oscillation experiment},''
  \bibinfo{journal}{Phys. Rev.} \textbf{\bibinfo{volume}{D64}},
  \bibinfo{pages}{112001} (\bibinfo{year}{2001}), \eprint{hep-ex/0107009}.

\bibitem[{\citenamefont{Gratta}()}]{GGratta}
\bibinfo{author}{\bibfnamefont{G.}~\bibnamefont{Gratta}},
  \bibinfo{howpublished}{Private Communication}.

\bibitem[{\citenamefont{An et~al.}(2017{\natexlab{a}})}]{An:2017osx}
\bibinfo{author}{\bibfnamefont{F.~P.} \bibnamefont{An}} \bibnamefont{et~al.}
  (\bibinfo{collaboration}{Daya Bay}), ``{Evolution of the Reactor Antineutrino
  Flux and Spectrum at Daya Bay},'' \bibinfo{journal}{Phys. Rev. Lett.}
  \textbf{\bibinfo{volume}{118}}, \bibinfo{pages}{251801}
  (\bibinfo{year}{2017}{\natexlab{a}}), \eprint{1704.01082}.

\bibitem[{\citenamefont{Bak et~al.}(2018{\natexlab{b}})}]{RENO:2018pwo}
\bibinfo{author}{\bibfnamefont{G.}~\bibnamefont{Bak}} \bibnamefont{et~al.}
  (\bibinfo{collaboration}{RENO}), ``{Fuel-composition dependent reactor
  antineutrino yield and spectrum at RENO},''
  (\bibinfo{year}{2018}{\natexlab{b}}), \eprint{1806.00574}.

\bibitem[{\citenamefont{Kim}()}]{SBKim}
\bibinfo{author}{\bibfnamefont{S.-B.} \bibnamefont{Kim}},
  \bibinfo{howpublished}{Private Communication}.

\bibitem[{\citenamefont{Esteban et~al.}(2019)\citenamefont{Esteban,
  Gonzalez-Garcia, Hern{\'a}ndez-Cabezudo, Maltoni, and
  Schwetz}}]{Esteban:2018azc}
\bibinfo{author}{\bibfnamefont{I.}~\bibnamefont{Esteban}},
  \bibinfo{author}{\bibfnamefont{M.~C.} \bibnamefont{Gonzalez-Garcia}},
  \bibinfo{author}{\bibfnamefont{A.}~\bibnamefont{Hern{\'a}ndez-Cabezudo}},
  \bibinfo{author}{\bibfnamefont{M.}~\bibnamefont{Maltoni}}, \bibnamefont{and}
  \bibinfo{author}{\bibfnamefont{T.}~\bibnamefont{Schwetz}}, ``{Global analysis
  of three-flavour neutrino oscillations: synergies and tensions in the
  determination of $\theta_{23}, \delta_{CP}$, and the mass ordering},''
  \bibinfo{journal}{JHEP} \textbf{\bibinfo{volume}{01}}, \bibinfo{pages}{106}
  (\bibinfo{year}{2019}), \eprint{1811.05487}.

\bibitem[{\citenamefont{Ashenfelter et~al.}(2016)}]{Ashenfelter:2015uxt}
\bibinfo{author}{\bibfnamefont{J.}~\bibnamefont{Ashenfelter}}
  \bibnamefont{et~al.} (\bibinfo{collaboration}{PROSPECT}), ``{The PROSPECT
  Physics Program},'' \bibinfo{journal}{J. Phys.}
  \textbf{\bibinfo{volume}{G43}}, \bibinfo{pages}{113001}
  (\bibinfo{year}{2016}), \eprint{1512.02202}.

\bibitem[{\citenamefont{Gariazzo et~al.}(2017)\citenamefont{Gariazzo, Giunti,
  Laveder, and Li}}]{Gariazzo:2017fdh}
\bibinfo{author}{\bibfnamefont{S.}~\bibnamefont{Gariazzo}},
  \bibinfo{author}{\bibfnamefont{C.}~\bibnamefont{Giunti}},
  \bibinfo{author}{\bibfnamefont{M.}~\bibnamefont{Laveder}}, \bibnamefont{and}
  \bibinfo{author}{\bibfnamefont{Y.~F.} \bibnamefont{Li}}, ``{Updated Global
  3+1 Analysis of Short-BaseLine Neutrino Oscillations},''
  \bibinfo{journal}{JHEP} \textbf{\bibinfo{volume}{06}}, \bibinfo{pages}{135}
  (\bibinfo{year}{2017}), \eprint{1703.00860}.

\bibitem[{\citenamefont{Giunti et~al.}(2017)\citenamefont{Giunti, Ji, Laveder,
  Li, and Littlejohn}}]{Giunti:2017yid}
\bibinfo{author}{\bibfnamefont{C.}~\bibnamefont{Giunti}},
  \bibinfo{author}{\bibfnamefont{X.~P.} \bibnamefont{Ji}},
  \bibinfo{author}{\bibfnamefont{M.}~\bibnamefont{Laveder}},
  \bibinfo{author}{\bibfnamefont{Y.~F.} \bibnamefont{Li}}, \bibnamefont{and}
  \bibinfo{author}{\bibfnamefont{B.~R.} \bibnamefont{Littlejohn}}, ``{Reactor
  Fuel Fraction Information on the Antineutrino Anomaly},''
  \bibinfo{journal}{JHEP} \textbf{\bibinfo{volume}{10}}, \bibinfo{pages}{143}
  (\bibinfo{year}{2017}), \eprint{1708.01133}.

\bibitem[{\citenamefont{Dentler et~al.}(2017)\citenamefont{Dentler,
  Hern\'{a}ndez-Cabezudo, Kopp, Maltoni, and Schwetz}}]{Dentler:2017tkw}
\bibinfo{author}{\bibfnamefont{M.}~\bibnamefont{Dentler}},
  \bibinfo{author}{\bibfnamefont{A.}~\bibnamefont{Hern\'{a}ndez-Cabezudo}},
  \bibinfo{author}{\bibfnamefont{J.}~\bibnamefont{Kopp}},
  \bibinfo{author}{\bibfnamefont{M.}~\bibnamefont{Maltoni}}, \bibnamefont{and}
  \bibinfo{author}{\bibfnamefont{T.}~\bibnamefont{Schwetz}}, ``{Sterile
  neutrinos or flux uncertainties? -- Status of the reactor anti-neutrino
  anomaly},'' \bibinfo{journal}{JHEP} \textbf{\bibinfo{volume}{11}},
  \bibinfo{pages}{099} (\bibinfo{year}{2017}), \eprint{1709.04294}.

\bibitem[{\citenamefont{Giunti et~al.}(2019)\citenamefont{Giunti, Li,
  Littlejohn, and Surukuchi}}]{Giunti:2019qlt}
\bibinfo{author}{\bibfnamefont{C.}~\bibnamefont{Giunti}},
  \bibinfo{author}{\bibfnamefont{Y.~F.} \bibnamefont{Li}},
  \bibinfo{author}{\bibfnamefont{B.~R.} \bibnamefont{Littlejohn}},
  \bibnamefont{and} \bibinfo{author}{\bibfnamefont{P.~T.}
  \bibnamefont{Surukuchi}}, ``{Diagnosing the Reactor Antineutrino Anomaly with
  Global Antineutrino Flux Data},'' \bibinfo{journal}{Phys. Rev.}
  \textbf{\bibinfo{volume}{D99}}, \bibinfo{pages}{073005}
  (\bibinfo{year}{2019}), \eprint{1901.01807}.

\bibitem[{\citenamefont{Hayes et~al.}(2014)\citenamefont{Hayes, Friar, Garvey,
  Jungman, and Jonkmans}}]{Hayes:2013wra}
\bibinfo{author}{\bibfnamefont{A.~C.} \bibnamefont{Hayes}},
  \bibinfo{author}{\bibfnamefont{J.~L.} \bibnamefont{Friar}},
  \bibinfo{author}{\bibfnamefont{G.~T.} \bibnamefont{Garvey}},
  \bibinfo{author}{\bibfnamefont{G.}~\bibnamefont{Jungman}}, \bibnamefont{and}
  \bibinfo{author}{\bibfnamefont{G.}~\bibnamefont{Jonkmans}}, ``{Systematic
  Uncertainties in the Analysis of the Reactor Neutrino Anomaly},''
  \bibinfo{journal}{Phys. Rev. Lett.} \textbf{\bibinfo{volume}{112}},
  \bibinfo{pages}{202501} (\bibinfo{year}{2014}), \eprint{1309.4146}.

\bibitem[{\citenamefont{Alekseev et~al.}(2018)}]{Alekseev:2018efk}
\bibinfo{author}{\bibfnamefont{I.}~\bibnamefont{Alekseev}} \bibnamefont{et~al.}
  (\bibinfo{collaboration}{DANSS}), ``{Search for sterile neutrinos at the
  DANSS experiment},'' \bibinfo{journal}{Phys. Lett.}
  \textbf{\bibinfo{volume}{B787}}, \bibinfo{pages}{56} (\bibinfo{year}{2018}),
  \eprint{1804.04046}.

\bibitem[{\citenamefont{Adey et~al.}(2018)}]{Adey:2018zwh}
\bibinfo{author}{\bibfnamefont{D.}~\bibnamefont{Adey}} \bibnamefont{et~al.}
  (\bibinfo{collaboration}{Daya Bay}), ``{Measurement of the Electron
  Antineutrino Oscillation with 1958 Days of Operation at Daya Bay},''
  \bibinfo{journal}{Phys. Rev. Lett.} \textbf{\bibinfo{volume}{121}},
  \bibinfo{pages}{241805} (\bibinfo{year}{2018}), \eprint{1809.02261}.

\bibitem[{\citenamefont{An et~al.}(2017{\natexlab{b}})}]{An:2016srz}
\bibinfo{author}{\bibfnamefont{F.~P.} \bibnamefont{An}} \bibnamefont{et~al.}
  (\bibinfo{collaboration}{Daya Bay}), ``{Improved Measurement of the Reactor
  Antineutrino Flux and Spectrum at Daya Bay},'' \bibinfo{journal}{Chin. Phys.}
  \textbf{\bibinfo{volume}{C41}}, \bibinfo{pages}{013002}
  (\bibinfo{year}{2017}{\natexlab{b}}), \eprint{1607.05378}.

\bibitem[{\citenamefont{Ashenfelter et~al.}(2018)}]{Ashenfelter:2018iov}
\bibinfo{author}{\bibfnamefont{J.}~\bibnamefont{Ashenfelter}}
  \bibnamefont{et~al.} (\bibinfo{collaboration}{PROSPECT}), ``{First search for
  short-baseline neutrino oscillations at HFIR with PROSPECT},''
  \bibinfo{journal}{Phys. Rev. Lett.} \textbf{\bibinfo{volume}{121}},
  \bibinfo{pages}{251802} (\bibinfo{year}{2018}), \eprint{1806.02784}.

\bibitem[{\citenamefont{Almaz\'{a}n et~al.}(2018)}]{Almazan:2018wln}
\bibinfo{author}{\bibfnamefont{H.}~\bibnamefont{Almaz\'{a}n}}
  \bibnamefont{et~al.} (\bibinfo{collaboration}{STEREO}), ``{Sterile Neutrino
  Constraints from the STEREO Experiment with 66 Days of Reactor-On Data},''
  \bibinfo{journal}{Phys. Rev. Lett.} \textbf{\bibinfo{volume}{121}},
  \bibinfo{pages}{161801} (\bibinfo{year}{2018}), \eprint{1806.02096}.

\bibitem[{\citenamefont{Bernard}(2019)}]{Bernard:2019jli}
\bibinfo{author}{\bibfnamefont{L.}~\bibnamefont{Bernard}}
  (\bibinfo{collaboration}{STEREO}), in \emph{\bibinfo{booktitle}{{54th
  Rencontres de Moriond on Electroweak Interactions and Unified Theories
  (Moriond EW 2019) La Thuile, Italy, March 16-23, 2019}}}
  (\bibinfo{year}{2019}), \eprint{1905.11896}.

\bibitem[{\citenamefont{Agostini and Neumair}(2019)}]{Agostini:2019jup}
\bibinfo{author}{\bibfnamefont{M.}~\bibnamefont{Agostini}} \bibnamefont{and}
  \bibinfo{author}{\bibfnamefont{B.}~\bibnamefont{Neumair}}, ``{Statistical
  Methods for the Search of Sterile Neutrinos},''  (\bibinfo{year}{2019}),
  \eprint{1906.11854}.

\bibitem[{\citenamefont{Alonso and Nakamura}(2017)}]{Alonso:2017fci}
\bibinfo{author}{\bibfnamefont{J.~R.} \bibnamefont{Alonso}} \bibnamefont{and}
  \bibinfo{author}{\bibfnamefont{K.}~\bibnamefont{Nakamura}}
  (\bibinfo{collaboration}{IsoDAR}), ``{IsoDAR@KamLAND: A Conceptual Design
  Report for the Conventional Facilities},''  (\bibinfo{year}{2017}),
  \eprint{1710.09325}.

\bibitem[{\citenamefont{Wong}(2008)}]{Wong:2008vk}
\bibinfo{author}{\bibfnamefont{H.~T.} \bibnamefont{Wong}}, ``{Ultra-Low-Energy
  Germanium Detector for Neutrino-Nucleus Coherent Scattering and Dark Matter
  Searches},'' \bibinfo{journal}{Mod. Phys. Lett.}
  \textbf{\bibinfo{volume}{A23}}, \bibinfo{pages}{1431} (\bibinfo{year}{2008}),
  \eprint{0803.0033}.

\bibitem[{\citenamefont{Akimov et~al.}(2013)}]{Akimov:2012aya}
\bibinfo{author}{\bibfnamefont{D.~{\relax Yu}.} \bibnamefont{Akimov}}
  \bibnamefont{et~al.} (\bibinfo{collaboration}{RED}), ``{Prospects for
  observation of neutrino-nuclear neutral current coherent scattering with
  two-phase Xenon emission detector},'' \bibinfo{journal}{JINST}
  \textbf{\bibinfo{volume}{8}}, \bibinfo{pages}{P10023} (\bibinfo{year}{2013}),
  \eprint{1212.1938}.

\bibitem[{\citenamefont{G\'{u}tlein et~al.}(2015)}]{Gutlein:2014gma}
\bibinfo{author}{\bibfnamefont{A.}~\bibnamefont{G\'{u}tlein}}
  \bibnamefont{et~al.}, ``{Impact of coherent neutrino nucleus scattering on
  direct dark matter searches based on CaWO$_4$ crystals},''
  \bibinfo{journal}{Astropart. Phys.} \textbf{\bibinfo{volume}{69}},
  \bibinfo{pages}{44} (\bibinfo{year}{2015}), \eprint{1408.2357}.

\bibitem[{\citenamefont{Belov et~al.}(2015)}]{Belov:2015ufh}
\bibinfo{author}{\bibfnamefont{V.}~\bibnamefont{Belov}} \bibnamefont{et~al.},
  ``{The $\nu$GeN experiment at the Kalinin Nuclear Power Plant},''
  \bibinfo{journal}{JINST} \textbf{\bibinfo{volume}{10}},
  \bibinfo{pages}{P12011} (\bibinfo{year}{2015}).

\bibitem[{\citenamefont{Aguilar-Arevalo
  et~al.}(2016)}]{Aguilar-Arevalo:2016qen}
\bibinfo{author}{\bibfnamefont{A.}~\bibnamefont{Aguilar-Arevalo}}
  \bibnamefont{et~al.} (\bibinfo{collaboration}{CONNIE}), ``{Results of the
  engineering run of the Coherent Neutrino Nucleus Interaction Experiment
  (CONNIE)},'' \bibinfo{journal}{JINST} \textbf{\bibinfo{volume}{11}},
  \bibinfo{pages}{P07024} (\bibinfo{year}{2016}), \eprint{1604.01343}.

\bibitem[{\citenamefont{Agnolet et~al.}(2017)}]{Agnolet:2016zir}
\bibinfo{author}{\bibfnamefont{G.}~\bibnamefont{Agnolet}} \bibnamefont{et~al.}
  (\bibinfo{collaboration}{MINER}), ``{Background Studies for the MINER
  Coherent Neutrino Scattering Reactor Experiment},'' \bibinfo{journal}{Nucl.
  Instrum. Meth.} \textbf{\bibinfo{volume}{A853}}, \bibinfo{pages}{53}
  (\bibinfo{year}{2017}), \eprint{1609.02066}.

\bibitem[{\citenamefont{Billard et~al.}(2017)}]{Billard:2016giu}
\bibinfo{author}{\bibfnamefont{J.}~\bibnamefont{Billard}} \bibnamefont{et~al.},
  ``{Coherent Neutrino Scattering with Low Temperature Bolometers at Chooz
  Reactor Complex},'' \bibinfo{journal}{J. Phys.}
  \textbf{\bibinfo{volume}{G44}}, \bibinfo{pages}{105101}
  (\bibinfo{year}{2017}), \eprint{1612.09035}.

\bibitem[{\citenamefont{Strauss et~al.}(2017)}]{Strauss:2017cuu}
\bibinfo{author}{\bibfnamefont{R.}~\bibnamefont{Strauss}} \bibnamefont{et~al.},
  ``{The $\nu$-cleus experiment: A gram-scale fiducial-volume cryogenic
  detector for the first detection of coherent neutrino-nucleus scattering},''
  \bibinfo{journal}{Eur. Phys. J.} \textbf{\bibinfo{volume}{C77}},
  \bibinfo{pages}{506} (\bibinfo{year}{2017}), \eprint{1704.04320}.

\bibitem[{\citenamefont{Akimov et~al.}(2017)}]{Akimov:2017ade}
\bibinfo{author}{\bibfnamefont{D.}~\bibnamefont{Akimov}} \bibnamefont{et~al.}
  (\bibinfo{collaboration}{COHERENT}), ``{Observation of Coherent Elastic
  Neutrino-Nucleus Scattering},'' \bibinfo{journal}{Science}
  (\bibinfo{year}{2017}), \eprint{1708.01294}.

\bibitem[{\citenamefont{Hakenm\"{u}ller}()}]{CONUStalk}
\bibinfo{author}{\bibfnamefont{J.}~\bibnamefont{Hakenm\"{u}ller}}, ``{The CONUS
  Experiment},'',
  \urlprefix\url{https://indico.cern.ch/event/606690/contributions/2591545/attachments/1499330/2336272/Taup2017_CONUS_talk_JHakenmueller.pdf}.

\bibitem[{\citenamefont{Ca{\~n}as et~al.}(2018)\citenamefont{Ca{\~n}as,
  Garc\'es, Miranda, and Parada}}]{Canas:2017umu}
\bibinfo{author}{\bibfnamefont{B.~C.} \bibnamefont{Ca{\~n}as}},
  \bibinfo{author}{\bibfnamefont{E.~A.} \bibnamefont{Garc\'es}},
  \bibinfo{author}{\bibfnamefont{O.~G.} \bibnamefont{Miranda}},
  \bibnamefont{and} \bibinfo{author}{\bibfnamefont{A.}~\bibnamefont{Parada}},
  ``{The reactor antineutrino anomaly and low energy threshold neutrino
  experiments},'' \bibinfo{journal}{Phys. Lett.}
  \textbf{\bibinfo{volume}{B776}}, \bibinfo{pages}{451} (\bibinfo{year}{2018}),
  \eprint{1708.09518}.

\bibitem[{\citenamefont{Blanco et~al.}(2019)\citenamefont{Blanco, Hooper, and
  Machado}}]{Blanco:2019vyp}
\bibinfo{author}{\bibfnamefont{C.}~\bibnamefont{Blanco}},
  \bibinfo{author}{\bibfnamefont{D.}~\bibnamefont{Hooper}}, \bibnamefont{and}
  \bibinfo{author}{\bibfnamefont{P.}~\bibnamefont{Machado}}, ``{Constraining
  Sterile Neutrino Interpretations of the LSND and MiniBooNE Anomalies with
  Coherent Neutrino Scattering Experiments},''  (\bibinfo{year}{2019}),
  \eprint{1901.08094}.

\end{thebibliography}

\end{document}